\documentclass[12pt]{article}
\usepackage{amsfonts}
\usepackage{amsmath}
\usepackage{amssymb}
\usepackage{amsthm}
\usepackage{geometry}
\usepackage{rotating}
\usepackage[utf8]{inputenc}
\usepackage{array}
\usepackage[english]{babel}
\usepackage{color}

\usepackage{bm}
\usepackage{natbib}

\usepackage{setspace}

\newcommand{\bfm}[1]   {\mbox{\boldmath{${#1}$}}}
\newcommand{\mean}   {\mbox{\textnormal{E}}}
\theoremstyle{plain} \newtheorem{proposition}{Proposition}

\begin{document}
\title{A high-dimensional two-sample test for the mean using random subspaces}

\author
{ M{\aa}ns Thulin\\{\footnotesize{Department of Mathematics, Uppsala University}}}
\date{}

\maketitle

% Start doublespacing
%\doublespacing

\begin{abstract}
\noindent A common problem in genetics is that of testing whether a set of highly dependent gene expressions differ between two populations, typically in a high-dimension\-al setting where the data dimension is larger than the sample size. Most high-dimensional tests for the equality of two mean vectors rely on naive diagonal or trace estimators of the covariance matrix, ignoring dependencies between variables. A test recently proposed by \citet{lo1} implicitly incorporates dependencies by using random pseudo-projections to a lower-dimensional space. Their test offers higher power when the variables are dependent, but lacks desirable invariance properties and relies on asymptotic p-values that are too conservative. We illustrate how a permutation approach can be used to obtain p-values for the Lopes et al. test and how modifying the test using random subspaces leads to a test statistic that is invariant under linear transformations of the marginal distributions. The resulting test does not rely on assumptions about normality or the structure of the covariance matrix. We show by simulation that the new test has higher power than competing tests in realistic settings motivated by microarray gene expression data. We also discuss the computational aspects of high-dimensional permutation tests and provide an efficient R implementation of the proposed test.

\noindent 
   \\[1.1mm] {\bf Keywords:} computational statistics; gene expression data; gene-set testing; high-dimensional data; large $p$ small $n$; permutation test; random subspace; test about the mean; two-sample problem.
\end{abstract}

%%%%%%%%%%%%%%%%%%%%%%%%%%%%%%%%%%%%%%%%%%%%%%%%%%%%%%%%%%%%%%%%%%%%
\section{Introduction}\label{introduction}
A commonly encountered problem in modern genetic research, geological imaging, signal processing, astrometry and finance is that of comparing the mean vectors of two populations. In many of today's applications, the data averts analysis by classic statistical methods as the data dimension $p$ typically is larger than the sample size $n$, which causes most standard procedures to break down. When $p<n$, comparisons of this type are usually done using Hotelling's $T^2$ test. In the high-dimensional setting where $p\geq n$, the sample covariance matrix is not invertible, meaning that Hotelling's test no longer can be used.

In this paper we discuss two-sample tests in a high-dimensional setting where the variables have a non-negligible dependence structure. While the tests are discussed in the context of gene-set testing, we stress that they are equally applicable to other fields in which high-dimensional data occur.

In genetic research, one is often interested in identifying differentially expressed genes between two groups of patients based on data from a microarray experiment. Genes, however, do not function in isolation. Rather, they work together in complex networks. It is therefore often of greater interest to search for sets of genes, rather than individual genes, that are differentially expressed \citep{ba1,ef2,gb1,ne1,re1,ba2,ch1}. These gene-sets are determined a priori, typically by utilizing databases such as Gene Ontology\footnote{http://www.geneontology.org/} or Kyoto Encyclopedia of Genes and Genomes\footnote{http://www.genome.jp/kegg/} or by grouping genes with similar chromosomal locations together.

A common approach to finding differentially expressed gene-sets is to use a two-step procedure, starting by performing individual tests for each gene. These gene-level tests are then aggregated into a single test for the entire gene set. Much, if not all, of the multivariate structure of the data set is lost when gene-level test are used. \citet{gb1}, \citet{ef1} and \citet{ga1} demonstrated several problems with common tests based on this approach, including very high rates of false positives. The empirical Bayes methods applied e.g. by \citet{ef3} suffer from similar issues, caused by correlations between gene expressions \citep{qui1}.

\begin{figure}
\begin{center}
 \caption{Comparison of two bivariate data sets. The red circles have population mean vector $(1,1)$ whereas the blue crosses have population mean vector $(0.25,1.75)$. The shifts are difficult to detect by looking at the marginal distributions, but become evident when comparing the joint distributions.}\label{fig0}
   \includegraphics[width=\textwidth]{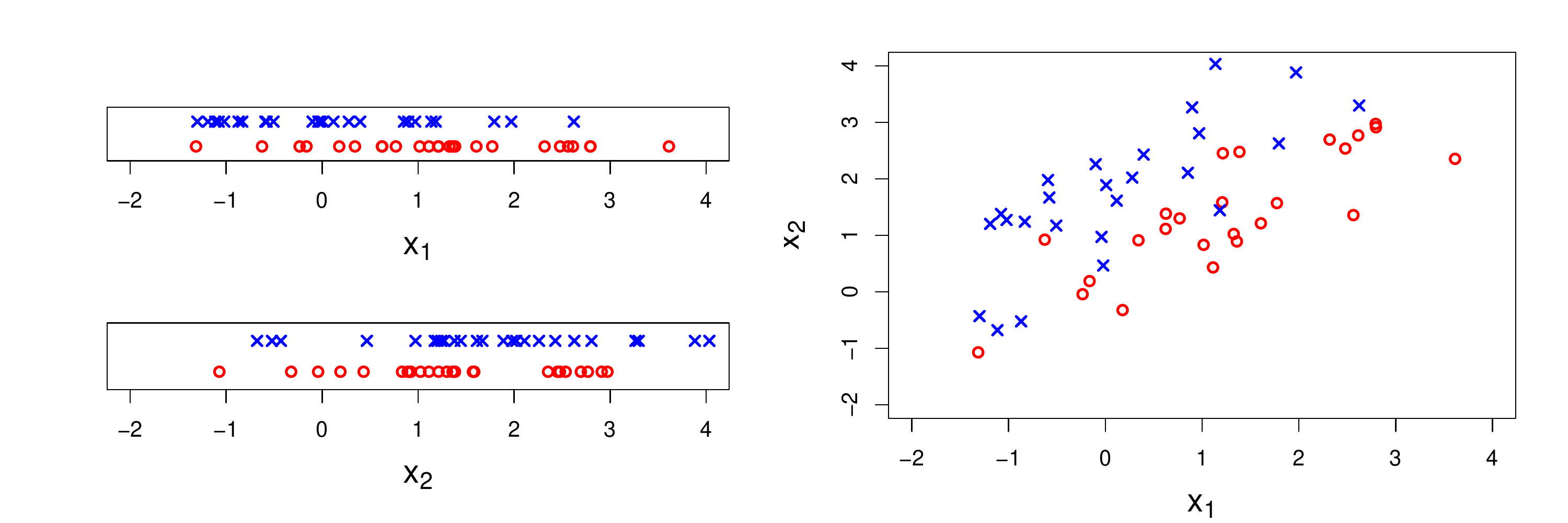}
 \end{center}
\end{figure}

By using a truly multivariate test for the gene set, it is possible not only to take the multivariate dependence structure of the gene expressions into account, but to gain more power from these dependencies, as illustrated in Figure \ref{fig0}.
There are three approaches to modifying the covariance estimator in Hotelling's test statistic to allow for high-dimensional inference. The first approach is to use prior information about the covariance structure to estimate the covariance matrix. \citet{ja1} presented such a test in the setting where location shift between the two populations is related to a known graph structure describing the dependence between the genes.

The second approach is to assume that the covariance matrix has a simple diagonal structure. The tests proposed by \citet{bs1}, \citet{sr1}, \citet{sr2}, \citet{ch1}, \citet{sr4} rely on imposing this particular structure on the covariance matrix, assuming the expressions of different genes to be independent. This is an unrealistic assumption for gene expressions, where genetic regulatory networks tend to cause the expression to be highly correlated. The assumption is equally unrealistic in many other biomedical problems. As a further example, the prevalence of an allele is typically highly correlated with the prevalence of other alleles on neighbouring loci.

The third approach is to use an estimator that allows for dependence, but that can be used in the absence of prior information. Recently, \citet{lo1} proposed a test in which the data is randomly pseudo-projected into several lower-dimensional spaces. Hotelling's $T^2$ statistic is computed for each pseudo-projection, and the result is then averaged over all pseudo-projections. Lopes et al. showed by asymptotic arguments and a simulation study that their test has substantially higher power than competing tests when the variables are correlated. There are however two downsides to their proposed method. First, it relies on an asymptotic null distribution derived under the assumption of normality. For finite sample sizes, this often leads to far too conservative p-values. Second, the test statistic is not invariant under linear transformations of the marginal distributions. This is a serious drawback, as it is common for genetic data to be rescaled by dividing the marginal distributions by their respective standard deviations.

In this paper, we show how accurate p-values for the Lopes et al. test can be obtained by using random permutations. We then propose a modified test statistic, which uses random subspaces instead of random pseudo-projections. The new test statistic is, conditioned on the random subspaces chosen, invariant under linear transformations of the marginal distributions.

A common problem in gene-set testing is that of identifying gene-sets, or pathways, that are related to cancer. For a pathway to induce cancer, a mutation must have occurred in at least one of its genes. Depending on where in this pathway the gene is located, the mutation can cause changes in the expressions of only a handful of genes or in all genes in the pathway. Motivated by this problem setting, we perform a Monte Carlo comparison of four multivariate two-sample tests and two tests based on gene-level $t$-tests. We compare the type I error rates and powers of the tests under different models for pathway dependencies and mutation locations. Some tests that require normality are modified so that p-values are computed using permutations rather than asymptotic null distributions, resulting in better type I error rates as well as higher power. We also contrast the invariance properties of the test statistics. While invariance properties tend to be overlooked in the biomedical literature, they are of great importance in multivariate testing and need to be taken into account when choosing which test to use.

High-dimensional permutation tests are heavily computer-intensive. For that reason, we discuss some computational aspects of such tests, and show how to efficiently implement the proposed test in R.

Methods based on random projections and random subspaces have not been studied to a great extent in the statistical literature, but are common in machine learning, where these techniques mainly have been been used for clustering and classification. See \citet{da1} and \citet{bm1} for reviews and some applications and \citet{vfl1} for applications in chemometrics. Most authors have used only a single random projection or subspace, although there are a few exceptions, including the the recent paper by \citet{lo1}. \citet{cafr1} used multiple random projections for goodness-of-fit testing but did not find the increase in power to be large enough to motivate the added computational complexity. \citet{zfb1} used multiple random projections for clustering, in a manner that bears resemblance to the algorithms presented in this paper, and found that it improved the performance of their clustering algorithms. Recently, \cite{isi1} proposed a subspace method for outlier detection in high-dimensional data sets that is somewhat similar to the random subspaces test presented in the present paper. Their method differs from ours in several ways, the most important difference being that their goal is to give an anomaly score \emph{to each observation}, rather than to compute a statistic that can be used for inference about the underlying population. Also worth mentioning is a recent paper by \citet{wei1}, who described a general hypothesis testing framework based on a non-random projection to the real line, determined by a linear classifier.

The rest of the paper is organised as follows. In Section \ref{projection} we review the Lopes et al. test and discuss its drawbacks. In Section \ref{projection2} we propose a new test based on random subspaces. In Section \ref{comp} we compare several gene-set tests in terms of invariance, type I error rates and power. In Section \ref{comput} we discuss the computational aspects of the new test. The text concludes with a discussion in Section \ref{disc} and an appendix with implementations and examples in R.

%%%%%%%%%%%%%%%%%%%%%%%%%%%%%%%%%%%%%%%%%%%%%%%%%%%%%%%%
\section{The Lopes et al. test}\label{projection}
\subsection{Setting}\label{pro00}
We consider two random samples of size $n_X$ and $n_Y$ from independent $p$-dimensional random variables $\bfm{X}$ and $\bfm{Y}$, with mean vectors $\bfm{\mu}_X$ and $\bfm{\mu}_Y$ and covariance matrices $\bfm{\Sigma}_X=\bfm{\Sigma}_Y=\bfm{\Sigma}$. We assume that $n_X+n_Y-2\geq p$. Our aim is to test whether $\bfm{\mu}_X=\bfm{\mu}_Y$.

One could argue that when testing the hypothesis $\bfm{\mu}_X=\bfm{\mu}_Y$ for two groups of patients, say a control group and a group of cancer patients, we should not expect the covariance matrices of the two groups to coincide. \citet{sr4} proposed a solution to this high-dimensional Behrens--Fisher problem. The hypothesis that we wish to test using mean vectors is however often not strictly speaking $\bfm{\mu}_X=\bfm{\mu}_Y$, but rather that the two multivariate distributions agree: $\bfm{F}_X=\bfm{F}_Y$. If the genes under consideration have no connection to cancer, there is no reason to expect the covariance matrix for their expressions to be any different from that of the control group. Under the null hypothesis that the two distributions are equal we should therefore assume that $\bfm{\Sigma}_X=\bfm{\Sigma}_Y$.

%%%%%%%%%%

\subsection{The test statistic}\label{pro0}
\citet{lo1} proposed a two-sample test of $\bfm{\mu}_X=\bfm{\mu}_Y$ that uses random projections to $k$-dimensional subspaces. Formally, assume that samples $ \bfm{X}=(\bfm{X}_1,\ldots,\bfm{X}_{n_X)}$ and $\bfm{Y}=(\bfm{Y}_1,\ldots,\bfm{Y}_{n_Y})$ are given. For $k\leq n_X+n_Y-2$, let $\bfm{P}_k$ be a $k\times p$ random matrix with i.i.d. $N(0,1)$ elements, independent of the data. 

The Lopes et al. statistic is computed by averaging the Hotelling's $T^2$ statistics over several random projections $\bfm{P}_k'\bfm{X}_1,\ldots,\bfm{P}_k'\bfm{Y}_{n_Y}$. The algorithm is as follows.
\noindent\begin{center}\framebox[\width][c]{
\begin{minipage}{\textwidth}
\subsubsection*{Algorithm 1: The Lopes et al. statistic $T_{L}$}
\begin{enumerate}
\item Generate a $k\times p$ random matrix $\bfm{P}_k$ with i.i.d. $N(0,1)$ elements. 
\item Calculate the test statistic $T_i^2$ based on the $n_X$ and $n_Y$ observations of the vectors $\bfm{X}_i=\bfm{P}_k'\bfm{X}$ and $\bfm{Y}_i=\bfm{P}_k'\bfm{Y}$.
\item Repeat steps 1-2 $B_1$ times, obtaining the test statistics $T^2_1,\ldots,T^2_{B_1}$.
\item Obtain the resulting test statistic as the function $T_{L}=B_1^{-1}\sum_{i=1}^{B_1}T_i^2$.
\end{enumerate}
\end{minipage}
}\end{center}
It should be noted that ``projection'' is used in a loose sense here, since $\bfm{P}_k$ usually isn't a projection matrix. Lopes et al. derived the asymptotic null distribution and power function of their test under normality and derived conditions under which the test is asymptotically more powerful than the tests of \citet{ch1} and \citet{sr2}.

In the following sections, we outline two drawbacks to this test and how the test procedure can be modified to account for these.

%%%%%%%%%%%%%%%%%%%%%%%%%%%%%%%%%%%%%%%%%

\subsection{The null distribution}\label{perm1}
\begin{figure}
\begin{center}
 \caption{Histograms of $T_{L}$ for 500 samples simulated under the null hypothesis. The asymptotic null distribution is shown in red.}\label{fig2}
   \includegraphics[width=\textwidth]{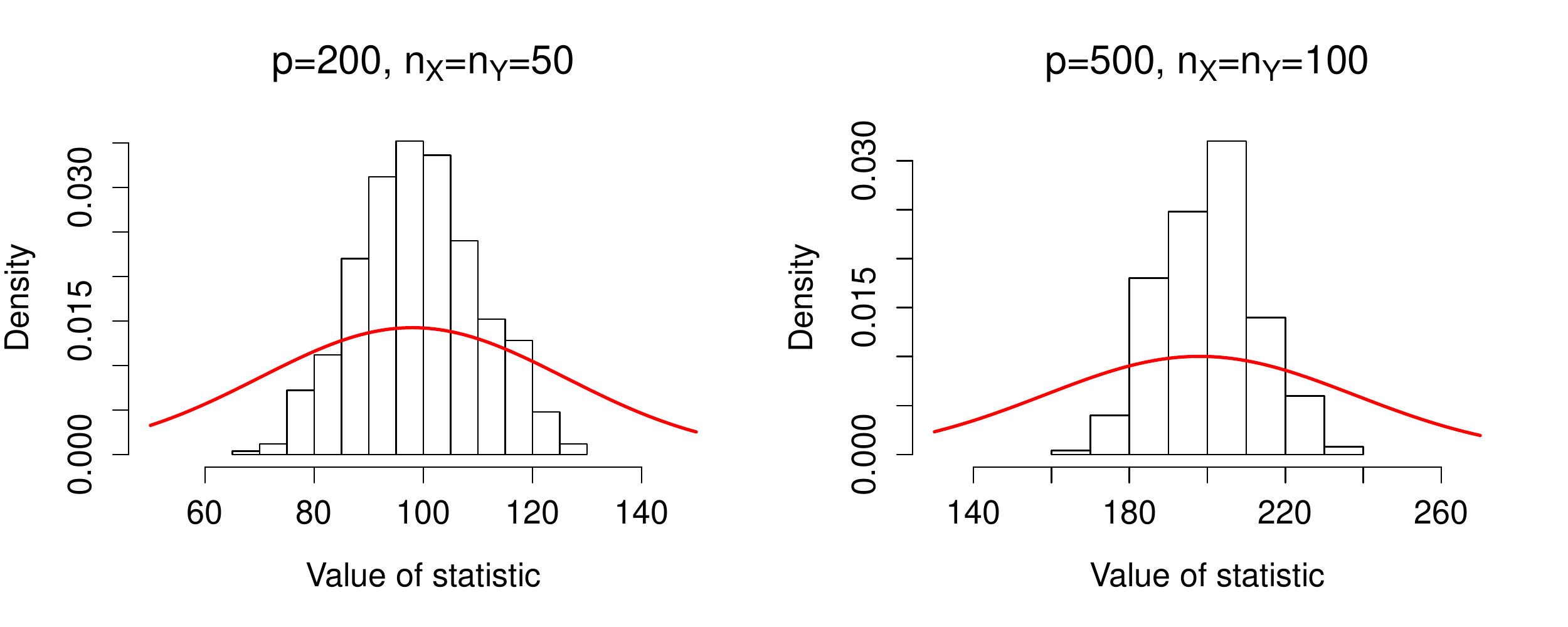}
   % Figur ritad med lopes2.R
 \end{center}
\end{figure}

The first drawback is that the asymptotic null distribution of the test statistic is a poor approximation of the null distribution for small sample sizes, such as the $p=200$, $n_X=n_Y=50$ setting studied in Section \ref{comp}. Two examples of this phenomenon are given in Figure \ref{fig2}. The p-values obtained from the asymptotic null distribution are highly conservative. For this reason, we instead propose using the permutation distribution of the data to compute the p-values.
Let $\bfm{Z}=(\bfm{Z}_1,\ldots,\bfm{Z}_{n_X+n_Y})=(\bfm{X}_1,\ldots,\bfm{X}_{n_X},\bfm{Y}_1,\ldots,\bfm{Y}_{n_Y})$ denote the entire sample. The number of permutations is usually too large for it to be computationally feasible to use the exact permutation distribution of $\bfm{Z}$. We can however obtain approximate p-values by using random permutations to approximate the permutation distribution. The algorithm is as follows.
\noindent\begin{center}\framebox[\width][c]{
\begin{minipage}{\textwidth}
\subsubsection*{Algorithm 2: A random permutation test}
\begin{enumerate}
\item Given a test statistic $T$, compute $T_{obs}=T(\bfm{X},\bfm{Y})$.
\item Draw $n_X$ integers $i_1,\ldots,i_{n_X}$ from $1,2,\ldots,n_X+n_Y$ without replacement. Let $j_1,\ldots,j_{n_Y}$ denote the numbers that were not chosen.
\item Calculate the test statistic $T$ for $\bfm{X}^*=(\bfm{Z}_{i_1},\bfm{Z}_{i_2},\ldots,\bfm{Z}_{i_{n_X}})$ and $\bfm{Y}^*=(\bfm{Z}_{j_1},\bfm{Z}_{j_2},\ldots,\bfm{Z}_{j_{n_Y}})$.
\item Repeat steps 1-2 $B_2$ times, obtaining the test statistics $T_{1},\ldots,T_{B_2}$.
\item Compute the p-value of the test as $B_2^{-1}\sum_{i=1}^{B_2}\mathbb{I}(T_{i}\geq T_{obs})$.
\end{enumerate}
\end{minipage}
}\end{center}

%%%%%%%%%%%%%%%%%%%%%%%%%%%%

\subsection{Invariance properties}\label{pro05}
The second drawback is that the statistic $T_{L}$ lacks desirable invariance properties. Ideally, the result of a statistical test should not depend on the location and scale on which the measurements have been obtained. Unfortunately, when $p>n_X+n_Y-2$, no two-sample test statistic that is a function of the sample can be affine invariant \citep[p. 318]{leh1}, i.e. invariant under all linear transformations. It is however possible for a test statistic to be invariant under a smaller group of linear transformations. 

In order to facilitate interpretability, gene expression data is often rescaled so that all variables have standard deviation 1, meaning that $\bfm{\Sigma}$ is a correlation matrix. This rescaling is a linear transformation of the marginal distributions, $\bfm{X}\to \bfm{DX}$, with $\bfm{D}$ being a diagonal matrix with diagonal elements $1/\sigma_i$, where $\sigma_i$ is the $i$:th standard deviation. In practice these standard deviations are almost invariably estimated. Invariance under such linear transformations of the marginal distributions is arguably the most important invariance property for a high-dimensional gene-set test.

$T_{L}$ is not invariant under this kind of linear transformations, even if we condition on the random projections $\bfm{P}_k$. Some R code that illustrates this is given in Appendix \ref{appinvar}. Using the data given in the appendix, for which $p=20$ and $n_X=n_Y=5$, we obtained the p-value $0.381$ when computing $T_{L}$ for the raw data with $k=4$ and $B_1=B_2=1000$. We then standardized the data by dividing by the marginal sample standard deviations. Keeping the random projections and permutations fixed, the p-value was $0.003$ after standardization. For the Lopes et al. test, standardization can turn a non-significant gene-set into a significant one.

In Section \ref{projection2} we propose a modification of the Lopes et al. test that leads to a test statistic that is invariant under linear transformations of the marginal distributions. The invariance properties of several two-sample test statistics are compared in Section \ref{comp1}.

%%%%%%%%%%%%%%%%%%%%%%%%%%%%%%%%%%%%%%%%%%%%%%%%%%%%%%%%
\section{A random subspaces test}\label{projection2}
%%%%%%%%%%%%%%%%

\subsection{Motivation and test procedure}\label{pro1}

Now, let $\bfm{X}_{(i)}$ and $\bfm{Y}_{(i)}$, $i=1,2,\ldots,\binom{p}{k}$, denote the $k$-dimensional subvectors of $\bfm{X}$ and $\bfm{Y}$, with $k\leq n_X+n_Y-2$. If $\bfm{\mu}_X=\bfm{\mu}_Y$ then $\mean(\bfm{X}_{(i)})=\mean(\bfm{Y}_{(i)})$ for all $i$. Tests of the hypothesis $\bfm{\mu}_X=\bfm{\mu}_Y$ can therefore be based on tests of the hypotheses $\mean(\bfm{X}_{(i)})=\mean(\bfm{Y}_{(i)})$.

$\binom{p}{k}$ is usually extremely large in the high-dimensional setting and studying all $k$-dimensional subvectors of $\bfm{X}$ and $\bfm{Y}$ is not feasible. As with the Lopes et al. test, we can however randomly select $B_1$ $k$-dimensional subvectors and compute $T^2$ for each subvector, after which a conclusion about the $p$-dimensional hypothesis can be drawn by averaging the test statistics for the different subvectors. Random subspace methods have previously been successfully applied to machine learning problems, e.g. by \citet{bert1} and \citet{carmen1}.

The heuristic motivation for this procedure is that if the deviations from the null hypothesis are spread out over many variables, they will likely be detected in many subvectors. On the other hand, if the deviations are concentrated in just a few variables, those variables are likely to be investigated if $B$ is large enough.

The algorithm for computing the random subspaces test statistic, denoted $T_{rs}$, is described below. In the rest of the paper the test statistic $T_i^2$ will be Hotelling's statistic, but the test can also be carried out using other statistics.
\noindent\begin{center}\framebox[\width][c]{
\begin{minipage}{\textwidth}
\subsubsection*{Algorithm 3: Computing the random subspace statistic $T_{rs}$}
\begin{enumerate}
\item Draw $k\leq n_X+n_Y-2$ integers $i_1,\ldots,i_k$ from $1,2,\ldots,p$ without replacement.
\item Calculate the test statistic $T_i^2$ based on the $n_X$ and $n_Y$  observations of the subvectors $\bfm{X}_{(i)}^*$ and $\bfm{Y}_{(i)}^*$, where the subvector with index $(i)$ is in the subspace consisting of the dimensions $i_1,\ldots,i_k$.
\item Repeat steps 1-2 $B_1$ times, obtaining the test statistics $T^2_1,\ldots,T^2_{B_1}$.
\item Obtain the resulting test statistic as the function $T_{rs}=B_1^{-1}\sum_{i=1}^{B_1}T^2_i$.
\end{enumerate}
\end{minipage}
}\end{center}

To compute the p-value for the test, we propose using the permutation distribution, i.e. using Algorithm 2 with the statistic $T_{rs}$.

%%%%%%%%%%%%%%%%%

\subsection{Properties and relation to the Lopes et al. test}\label{pro2}

While Lopes et al. proposed using $\bfm{P}_k$ with i.i.d. $N(0,1)$ elements, in their theoretical investigations they studied a more general test procedure, where $\bfm{P}_k$ can be generated by some other distribution.

\begin{proposition}
$T_{rs}$ is a special case of the general random projections test studied in \citet{lo1}.
\end{proposition}

To see this, let $\bfm{1}_{i}=(j_1,j_2,\ldots,j_p)$ be a $p$-vector with $j_h=\mathbb{I}_i(h)$. If $i_1,\ldots,i_k$ are drawn uniformly at random from $\{1,2,\ldots,p\}$ without replacement and if $\bfm{P}_k=(\bfm{1}_{i_1},\bfm{1}_{i_2},\ldots,\bfm{1}_{i_k})$ then the projected sample is simply the part of the sample that resides in the $k$-dimensional subspace given by the indices $i_1,\ldots,i_k$. Thus the random subspaces test and the general random projections test coincide for this particular choice of $\bfm{P}_k$.

The asymptotic results of \citet{lo1} hold for any random matrices $\bfm{P}_k$ that have full rank with probability 1. Since the $\bfm{P}_k$ described above always has full rank, we have the following result.

\begin{proposition}
$T_{rs}$ and $T_{L}$ are asymptotically equivalent.
\end{proposition}

Consequently, the asymptotic relative efficiency of the random subspaces test with respect to the Chen--Qin test is given in Theorem 3 of \citet{lo1} and the asymptotic relative efficiency with respect to the Srivastava--Du test is given in Theorem 4 of said paper.

While asymptotically equivalent, $T_{L}$ and $T_{rs}$ differ in their finite-sample properties, in particular regarding invariance:
\begin{proposition}
Let $\bfm{D}$ be a diagonal $p\times p$ real matrix with nonzero diagonal elements and let $\bfm{d}\in\mathbb{R}^p$. Conditioned on the random subspaces chosen, $T_{rs}$ is invariant under the linear transformations $(\bfm{X},\bfm{Y})\to (\bfm{DX}+\bfm{d},\bfm{DY}+\bfm{d})$.
\end{proposition}

Unlike e.g. orthogonal transformations, linear transformations of this type are linear in all $k$-dimensional subspaces, so that the invariance of $T_{rs}$ follows from the fact that Hotelling's statistic is invariant under linear transformations in the $k$-dimensional subspaces.

%%%%%%%%%%%%%%%%%

\subsection{Choosing $k$, $B_1$ and $B_2$}\label{pro3}
The choice of $k$ affects the performance of the $T_{rs}$ test. If $k$ is too close to $n_X+n_Y-2$ the test loses power, since Hotelling's $T^2$ performs poorly in this setting \citep{bs1}. If $k$ is too small, much of the multivariate structure of the data set is lost.

\citet{lo1} found analytically that the asymptotic power of their test was maximized when $k=\lfloor (n_X+n_Y-2)/2 \rfloor$. We have verified numerically that this seems to be a good choice for $T_{rs}$ for finite sample sizes, although our simulation results indicate that the power is quite insensitive to small changes in $k$, as illustrated in Section \ref{comp3}. For large $n_X$ and $n_Y$, one may for purely computational reasons have to use a smaller $k$, as the permutation step may be computationally unfeasible when $k$ is too large.

The power of the test increases slightly with $B_1$ and $B_2$, but is relatively stable for $B_1,B_2\geq 100$. In the simulation study below, we used $B_1=100$ and $B_2=500$.

%%%%%%%%%%%%%%%%%%%%%%%%%%%%%%%%%%%%%%%%%%%%%%%%%%%%%%%%%%%
%%%%%%%%%%%%%%%%%%%%%%%%%%%%%%%%%%%%%%%%%%%%%%%%%%%%%%%%%%%

\section{Comparison of two-sample tests}\label{comp}

\subsection{Tests to be compared}\label{comp1}
In order to evaluate the performance of the random subspaces test and the random permutations version of the \citet{lo1} test, we compared them to other two-sample tests.

\citet{bs1} proposed a Hotelling-type test utilizing the trace of the sample covariance matrix. More recently, \citet{ch1} proposed a modification of the Bai--Saranadasa test, which has higher power in most situations. We therefore excluded the  Bai--Saranadasa test from the study, choosing instead to focus our attention on the Chen--Qin test. We decided to use the asymptotic null distribution of the Chen--Qin statistic in our simulation, since using random permutations for computing the Chen--Qin p-value was extremely computer-intensive, at least using our implementation of the test.

\citet{sr1} proposed using the Moore-Penrose inverse of the sample covariance matrix $\bfm{S}$ when computing Hotelling's test statistic. In a small pilot study, we found the Srivastava test to be computationally expensive and to have lower power than the competing tests. It was therefore excluded from the larger study.

\citet{sr2} proposed replacing the sample covariance matrix $\bfm{S}$ in Hotelling's statistic by a diagonal estimator. Their test statistic is asymptotically standard normal under certain conditions, but convergence to normality appears to be slow. We used random permutations to compute the p-values of the Srivastava--Du-test in our comparison. Compared to using the asymptotic null distribution, we found that the permutation procedure provided better type I error rates and resulted in higher power.

Finally, we considered test based on combined multiple $t$-tests. For such tests, the multivariate null hypothesis is rejected if at least one of the marginal null hypotheses is rejected. We used two methods for combining the tests: Bonferroni correction, controlling the family-wise error rate, and the \citet{bh1} procedure, controlling the false discovery rate.

%%%%%%%%%%%%%%%%%%%%%%%%%%%%%%%%%%%%%%%%%

\subsection{Invariance properties}\label{comp1}
When choosing between different tests, we are usually concerned with how well they attain their nominal type I error rates and how high their power is. In the high-dimensional setting, we must also take invariance properties into account, as discussed in Section \ref{pro05}. We assume that both samples are transformed analogously.

In Section \ref{pro05} we argued that invariance under linear transformations of the marginal distributions is the most desirable invariance property in the gene expression setting. These are transformations of the type $\bfm{X}\to \bfm{DX}+\bfm{d}$, where $\bfm{D}$ is a real diagonal $p\times p$ matrix with nonzero diagonal elements and $\bfm{d}\in\mathbb{R}^p$. Among the tests considered here, only the marginal $t$-tests, the Srivastava--Du test and the random subspaces test (conditioned on the random subspaces chosen) are invariant under such transformations.

The Bai--Saranadasa, Chen--Qin and Srivastava tests are invariant under orthogonal transformations, i.e. transformations $\bfm{X}\to c\bfm{HX}$, where $c$ is a nonzero constant and $\bfm{H}$ is a real $p\times p$ orthogonal matrix. This invariance property is arguably less attractive in the gene expression setting, as rotations of the coordinate system are of little interest.

The Lopes et al. test is only invariant if the same scaling is applied to all marginal distributions, that is, under transformations of the type $\bfm{X}\to c\bfm{IX}$, where $\bfm{I}$ is the $p\times p$ identity matrix and $c$ is a non-zero constant.

%%%%%%%%%%%%%%%

\subsection{Type I error rates}\label{comp2}
All tests under consideration are based on either asymptotic or random permutation estimates of the null distribution. They do therefore in general not attain the nominal type I error rates exactly. To evaluate the actual type I error rates for the tests in a few interesting settings, a Monte Carlo study was performed under the null hypothesis $\bfm{\mu}_X=\bfm{\mu}_Y$ for $p=200$ and $n_X=n_Y=50$.

Two families of distributions of $\bfm{X}$ and $\bfm{Y}$ and three different covariance structures were evaluated in the simulations. Let $\bfm{\Sigma}_{a,b}$ denote a covariance matrix with unit variances and 8 equal-sized blocks, where $\mbox{Cov}(X_i,X_j)$ is $a$ if $X_i$ and $X_j$ belong to the same block and $b$ otherwise. The first three distributions were multivariate normal, with covariance matrices $\bfm{\Sigma}_{0,0}$, $\bfm{\Sigma}_{0.5,0.1}$ and $\bfm{\Sigma}_{0.9,0.2}$. To study the impact of heavy-tailed distributions, the last two were multivariate $t$-distributions with 4 degrees of freedom and the covariance matrices $\bfm{\Sigma}_{0,0}$ and $\bfm{\Sigma}_{0.5,0.1}$.

For each distribution, $1,000$ samples were generated under the null hypothesis. The tests were then applied to each sample, using $k=49$, $B_1=100$ and $B_2=500$ for $T_{rs}$ and $T_{L}$. The point estimates of the tests' type I error rates are given along with 95 \% confidence intervals (not adjusted for multiplicity) in Table \ref{htabell3}. All Hotelling-type tests have acceptable type I error rates, whereas the multiple $t$-tests have too low rates in some circumstances.

\begin{table}[h]
\begin{center}
\caption{Type I error rate of two-sample tests when $p=200$, $n_X=n_Y=50$ and $\alpha=0.05$.}\label{htabell3}
\footnotesize{
\begin{tabular}{|l | c |c  |c|c| c| c | }
\hline
 &  Bonferroni  & Benjamini-- & Chen--Qin   & Srivastava-- & Lopes & Random \\
 & $t$& Hochberg $t$ & & Du & et al. $T_{L}$ & subspaces $T_{rs}$\\\hline
Normal  & 0.047 & 0.048 & 0.051  & 0.046 & 0.047 & 0.049\\
$\bfm{\Sigma}_{0,0}$ &  \tiny{(0.035,0.061)} & \tiny{(0.036,0.062)} & \tiny{(0.039,0.066)}  & \tiny{(0.034,0.060)} & \tiny{(0.035,0.061)} & \tiny{(0.037, 0.063)} \\\hline
Normal & 0.056 & 0.058 & 0.065  & 0.052 & 0.046 & 0.058\\
$\bfm{\Sigma}_{0.5,0.1}$ & \tiny{(0.043,0.072)} & \tiny{(0.45,0.074)} & \tiny{(0.051,0.081)} &  \tiny{(0.040, 0.067)} & \tiny{(0.034,0.060)} & \tiny{(0.045,0.074)}\\\hline
Normal & 0.018 & 0.022 & 0.063  & 0.049 & 0.048 & 0.052\\
$\bfm{\Sigma}_{0.9,0.2}$& \tiny{(0.011,0.028)} & \tiny{(0.014,0.033)} & \tiny{(0.049,0.079)} &  \tiny{(0.037,0.064)} & \tiny{(0.036,0.0.63)} & \tiny{(0.040,0.067)}  \\\hline
$t(4)$ & 0.028 & 0.028 &0.056  & 0.058 &0.046 &0.051\\
$\bfm{\Sigma}_{0,0}$& \tiny{(0.019,0.040)} & \tiny{(0.019,0.040)} & \tiny{(0.043,0.072)} &  \tiny{(0.045,0.074)} & \tiny{(0.034,0.06)} & \tiny{(0.039,0.066)}\\\hline
$t(4)$ & 0.036 & 0.039 & 0.085 &  0.071 & 0.059 & 0.066\\
$\bfm{\Sigma}_{0.5,0.1}$& \tiny{(0.026,0.049)} & \tiny{(0.028,0.052)} & \tiny{(0.069,0.103)} &  \tiny{(0.056,0.088)} & \tiny{(0.046,0.075)} & \tiny{(0.052,0.083)} \\\hline
\end{tabular}
}
\end{center}
\end{table}

%%%%%%%%%%%%%%%

\subsection{Power study}\label{comp3}
\begin{figure}
\begin{center}
 \caption{Power of two-sample tests when $p=200$, $n_X=n_Y=50$ and $\alpha=0.05$.}\label{fig1}
   \includegraphics[width=\textwidth]{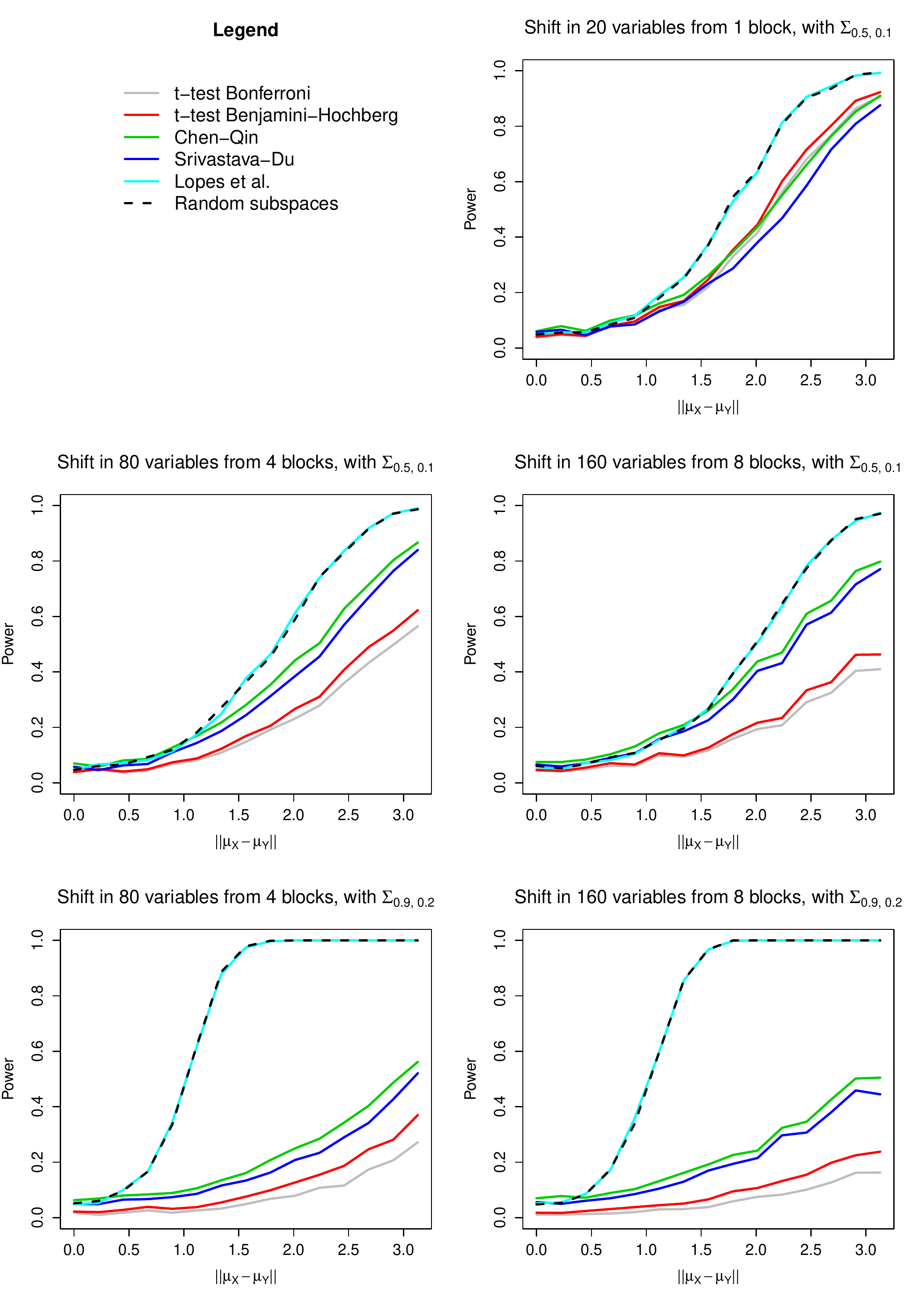}
 \end{center}
\end{figure}

To study the power of the tests in the cancer pathway setting described in Section \ref{introduction}, we performed simulations under the assumptions of normality and the covariance matrices $\bfm{\Sigma}_{0.5,0.1}$ and $\bfm{\Sigma}_{0.9,0.2}$ described in the previous section. For the $\bfm{Y}$ variable, we shifted the means of 20 out of 25 genes evenly in each of $m$ out of the 8 blocks, for $m\in\{1,4,8\}$. The powers of the tests as functions of the resulting Euclidean distance $||\bfm{\mu}_X-\bfm{\mu}_Y||$ are shown in Figure \ref{fig1}.

In the simulations, 1,000 samples were generated from each distribution. For the Lopes et al. and random subspaces tests, $B_1=100$, $B_2=500$ and $k=49$ were used. We also performed comparisons for other $p$, $n_X$ and $n_Y$, as well as settings where $\bfm{\Sigma}_X\neq\bfm{\Sigma}_Y$ and where the shifts were unevenly distributed among the variables. The resulting plots were not qualitatively different from those in Figure \ref{fig1} and are therefore not shown here.

In our comparison, the Lopes et al. test and the random subspaces tests were the only tests that \emph{gained} power as the correlation between the variables was increased. For all the other tests, the power became lower when the variables became more dependent. For a fixed covariance matrix, the power of the Hotelling-type tests are relatively stable when the number of variables in which there is a difference change. 
The multiple $t$-tests are much more sensitive to this type of changes: their power decreases as the number of shifted variables increases.

\begin{figure}[h]
\begin{center}
 \caption{Power of the random subspaces test for different $k$ when $p=200$, $n_X=n_Y=50$ and $\alpha=0.05$.}\label{fig3}
   \includegraphics[trim=0cm 20cm 0cm 0cm,clip=true,width=\textwidth]{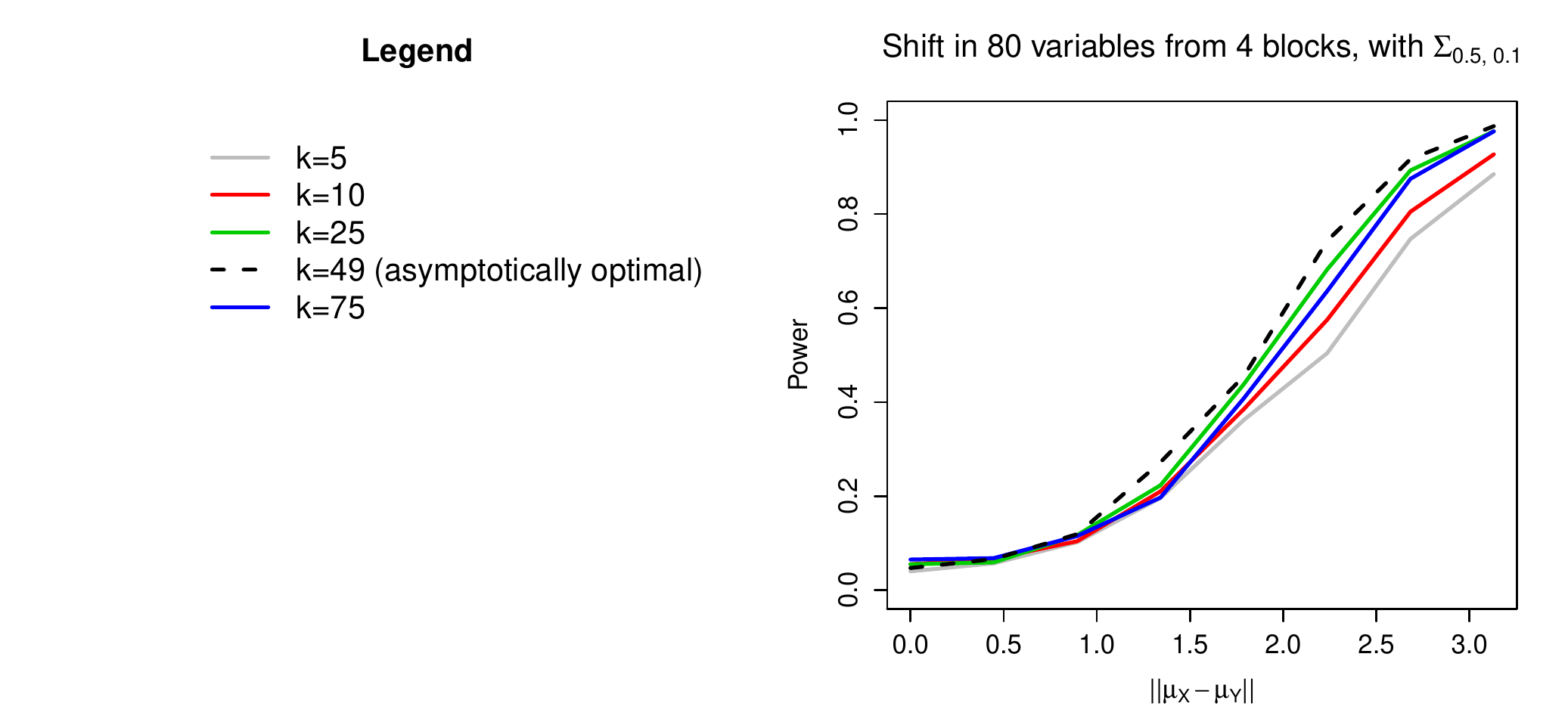}
 \end{center}
\end{figure}

In Figure \ref{fig3} we plot the power of $T_{rs}$ for different choices of $k$ for one of the alternatives from Figure \ref{fig1}. The plots for other alternatives are similar. While $k=\lfloor (n_X+n_Y-2)/2\rfloor$ gives the highest power, the test is surprisingly insensitive to changes in $k$. In the $p=200$, $n_X=n_Y=50$ setting, where $k=49$ is asymptotically optimal, there is little difference in power when $25\leq k \leq 75$. Under this particular alternative, the test is on a par with the Chen--Qin and Srivastava--Du tests even when $k=5$ (cf. Figure \ref{fig1}).

%%%%%%%%%%%%%%%%%%%%%%%%%%%%%%%%%%%%%%%%%%%%%%%%%%%%%%%%
\section{Computational aspects}\label{comput}
The random subspaces test quickly becomes very computer-intensive as $k$ increases, as we must invert $B_1\cdot B_2$ matrices of size $k\times k$ in order to obtain the p-value. In this section we discuss some ways to speed up the computations and compare the performances of different implementations in R.

Small gains in speed can be achieved by avoiding repeating the same calculations more than once, e.g. by computing $n=n_X+n_Y-2$ only once. Moreover, when using the permutation distribution instead of the asymptotic null distribution, there is for instance no need to rescale the test statistic by the sample size.

The most important tool for improving the computational speed, however, is parallelization, i.e. running the repetitions of steps 1-2 of Algorithm 2 simultaneously instead of sequentially. An efficient parallel R implementation of the random subspaces test is given in Appendix \ref{appR}.

The choice of $k$ impacts not only the power of the test, as shown in Figure \ref{fig3}, but also the computational cost of the algorithm. There is therefore a trade-off between computational speed and power; $k=\lfloor (n_X+n_Y-2)/2 \rfloor$ yields the highest power, but smaller $k$ will give improved computational performance. Similarly, larger $B_1$ and $B_2$ are preferable from a statistical viewpoint, but lead to increased computational costs.

\begin{figure}
\begin{center}
 \caption{Mean execution time of a single random subspaces test for different $k$ when $p=200$ and $n_X=n_Y=50$.}\label{fig4}
   \includegraphics[width=0.5\textwidth]{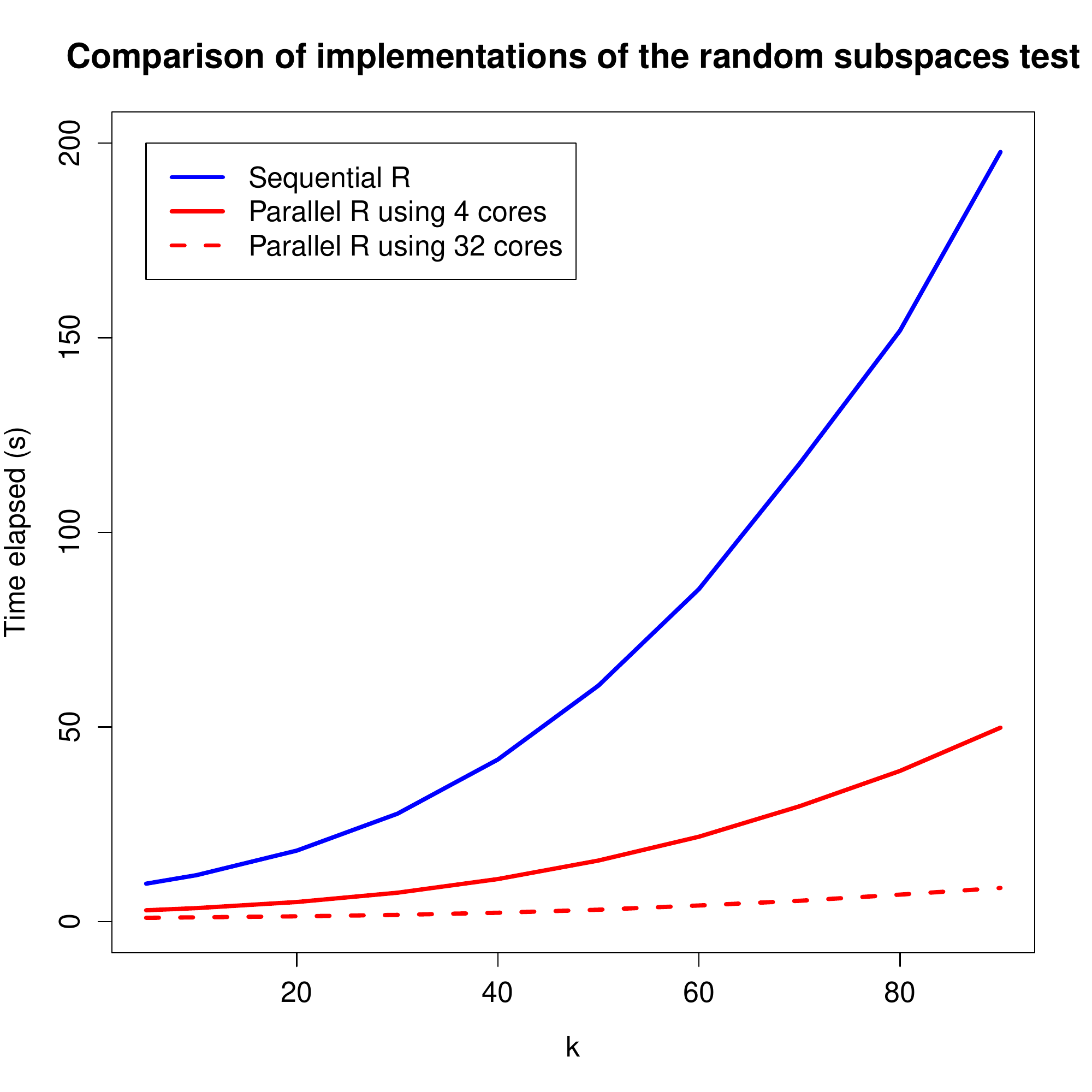}
 \end{center}
\end{figure}

In Figure \ref{fig4} the performances of two implementations of the test are compared for different choices of $k$ when $p=200$ and $n_X=n_Y=50$. The first is a naive sequential implementation in R. The second is a parallel implementation in R, using the \texttt{doMC} package. The implementations were executed on a 64-core 2.2 GHz AMD processor with 128 GB RAM and R 2.15.2, using different numbers of cores. The difference between the sequential and parallel R implementations is quite striking: when $k=50$, the sequential implementation needed 61 second to perform the test, whereas the parallel implementation only needed 15.7 seconds using 4 cores and 3.0 seconds using 32 cores.

%%%%%%%%%%%%%%%%%%%%%%%%%%%%%%%%%%%%%%%%%%%%%%%%%%%%%%%%
\section{Discussion}\label{disc}
\subsection{Recommendations for two-sample tests}\label{disc1}
Non-negligible dependence structures are present in virtually all genetic datasets and tests for differentially expressed gene-sets should therefore take such dependencies into account. Most high-dimensional two-sample tests, as well as multiple $t$-tests, do not do this to a sufficient degree. The Lopes et al. and random subspaces tests do account for dependencies, and consequently outperform their competitors in settings with non-negligible dependence structures. The powers of the two tests closely mimic each other. Unlike the Lopes et al. test however, the random subspaces test has the additional benefit of invariance under linear transformations of the marginals. Judging from the simulation results in Section \ref{comp}, we therefore recommend the random subspaces test as the default high-dimensional two-sample test.

%%%%%%%%%%%%%%%%%%%%%%%%%%%%%%%%%%%%%%%%%%%%%%%%%%%%%%%%
\subsection{Testing multiple gene-sets}\label{apps}
In most studies, one wishes to perform tests for a large number of gene-sets and not just for a single set. It is generally agreed upon that one should use some sort of adjustment for multiplicity in such investigations, in order to lower the number of false discoveries. Methods for doing such adjustments include Bonferroni corrections for controlling the family-wise error rate \citep[e.g.][]{ho1}, the \citet{bh1} false discovery rate procedure and resampling methods \citep{dud0,ge1,ba1,dud2,so1}. The latter class of methods seems especially promising when dependent gene-sets are tested, as resampling methods to a greater extent can account for such dependencies. When used with test procedures that use random permutations to compute p-values, such methods can however become extremely computer-intensive. It should also be noted that research on multiple testing in genetic research mostly has focused on gene-level $t$-tests and that recommendations for univariate test need not carry over to the multivariate setting. How to adjust high-dimensional tests for multiplicity remains an open problem.

%%%%%%%%%%%%%%%%%

\subsection*{Acknowledgments}
The author wishes to thank Elisabeth Thulin and Silvelyn Zwanzig for helpful discussions. Miles Lopes kindly provided R code for computing the Lopes et al. statistic, allowing the author to compare their implementation of the test to his own.

%%%%%%%%%%%%%%%%%%%%%%%%%%%%%%%%%%%%%%%%%%%%%%%%%%%%%%%%%%%%%%%%%%%%%%%%%%%%%%%%%%
%\pagebreak

~\\[3mm]
Contact information:\\
M{\aa}ns Thulin, Department of Mathematics, Uppsala University, Box 480, 751 06 Uppsala, Sweden\\
E-mail: thulin@math.uu.se

%%%%%%%%%%%%%%%%%%%%%%%%%%%%%%%%%%%%%%%%%%%%%%%%%%%%%%%%%%%%%%%%%%%%%%%%%%%%%%%%%%
\pagebreak
\appendix

\section{Appendix: An R implementation of the random subspaces test}\label{appR}
An implementation of the random subspaces statistic in R is given below. In order to speed up the computation of the statistic, the \texttt{.Internal} versions of \texttt{mean} and \texttt{cov} are used. This means that the function avoids error handling, such as checking whether the data contains \texttt{NA} values. The outer loop used in the computation is parallelized using the \texttt{foreach} and \texttt{doMC} packages.

\begin{verbatim}
# Load required packages:
library(compiler) # Compilation - for better performance
library(doMC)     # Parallelization - for better performance
registerDoMC()

# Hotellings T^2 statistic
T2.func<-function(x,y,n1,n2,p)
{
    mdiff<-.Internal(colMeans(x,n1,p,na.rm=FALSE))-
           .Internal(colMeans(y,n2,p,na.rm=FALSE))
    Spool<-((n1-1)*.Internal(cov(x, NULL, 1, FALSE))+(n2-1)*
                  .Internal(cov(y, NULL, 1, FALSE)))/(n1+n2-2)
    return(t(mdiff) %*% solve((1/n1+1/n2)*Spool) %*% mdiff)
}
T2.func<-cmpfun(T2.func)

# Random subspaces
subspacesT2<-function(x,y,n1,n2,p,B=100,k=floor((n1+n2-2)/2))
{
    res<-vector(length=B)
    for(j in 1:B)
    {
      x.cols<-sample(p,k)
      x.new<-x[,x.cols]
      y.new<-y[,x.cols]
      res[j]<-T2.func(x.new,y.new,n1,n2,k)
   }
   return(.Internal(mean(res)))
}
subspacesT2<-cmpfun(subspacesT2)

subspaces.test<-function(x,y,n1,n2,p,B1=100,B2=100,k=floor((n1+n2-2)/2))
{
    z<-rbind(x,y) # Big matrix to resample from
    zsize<-n1+n2
	
    # Permutations
    rs<-data.frame(foreach(i = 1:B1) %dopar%
    {
      x.rows<-sample(zsize,n1)
      x.new<-z[x.rows,]
      y.new<-z[-x.rows,]
      subspacesT2(x.new,y.new,n1,n2,p,B2,k)
    })
	
    rs.obs<-subspacesT2(x,y,n1,n2,p)

    # Return p-value:
    return(sum(rs>=as.numeric(rs.obs))/B1)
}
subspaces.test<-cmpfun(subspaces.test)

# Example of usage:
library(MASS)    # Used to generate multivariate normal data

# Set parameters:
p<-200
n1<-n2<-50
mu1<-rep(1,p)
mu2<-rep(1.02,p)
Sigma1<-matrix(0.25,p,p)
diag(Sigma1)<-1

# Generate example data:
x<-mvrnorm(n1, mu1, Sigma1)
y<-mvrnorm(n2, mu2, Sigma1)

# Apply test:
subspaces.test(x,y,n1,n2,p,B1=100,B2=500,k=49)
\end{verbatim}

%%%%%%%%%%%%%%%%%%%%%%%%%%%%%%%%%%%%%%%%%%%%%

\section{Appendix: Invariance under linear transformations of the marginals}\label{appinvar}
The example below illustrates the lack of invariance of the Lopes et al. test, discussed in Section \ref{pro05}. It uses the function \texttt{T2.func} from the previous section.

\begin{verbatim}
library(doMC)
registerDoMC()
library(MASS) 
library(compiler)

# Compute the random projections statistic
projectionT2<-function(x,y,n1,n2,p,B=100,k=floor((n1+n2-2)/2),matrixlist)
{
   res<-vector(length=B)
   for(j in 1:B)
   {
      P<-as.matrix(data.frame(matrixlist[j]))
      x.new<-x %*% P
      y.new<-y %*% P
      res[j]<-T2.func(x.new,y.new,n1,n2,k)
   }
   return(.Internal(mean(res)))
}
projectionT2<-cmpfun(projectionT2)

# Set parameters:
p<-20
n1<-5
n2<-5
k<-4
B<-100
B2<-500

# Import data:
x<-matrix(c(1.46,-2.28,0.73,0.02,0.39,0.75,-0.43,1.9,1.23,-0.15,1.31,2.37,
5.37,-2.03,-1.48,5.16,-2.72,-2.67,3.9,-0.97,0.84,-2.55,1.37,-1.53,1.7,
4.03,-0.1,0.97,4.24,0.43,3.13,-5.38,0.13,4.67,6.01,2.75,-1.71,3.52,2.17,
-2.93,2.45,2.59,-1.59,5.64,4.8,5.01,3.15,4.36,5.27,-0.53,1.58,0.53,1.39,
1.67,0.16,1.32,0.61,1.54,1.81,0.59,1.4,2.17,1.8,0.34,0.74,0.06,1.24,1.44,
0.91,0.55,0.22,1.32,0.36,0.94,1.34,1.87,0.69,0.65,1.62,0.16,0.28,-0.3,0.84,
1.28,1.33,2.3,1.55,1.54,1.87,1.29,2.2,0.9,1.44,2.02,1.34,1.73,1.92,0.31,
0.81,0.75),5,20)

y<-matrix(c(1.57,2.03,0.58,3.2,2.03,4.18,3.04,0.4,1.73,2.3,3.2,2.84,3.39,
4.17,2.84,1.26,2.88,-1.07,4.4,-0.71,-5.21,-2.07,2.7,6.02,-1.38,-0.03,3.06,
0.29,4.15,2.02,3.07,3.86,5.81,2.62,0,4.66,3.3,0.37,2.57,4.57,3.86,3.46,-1,
2.72,-1.58,2.06,6.09,6.88,1.36,0.07,2.33,-0.17,2.37,1.85,1.15,3.77,1.1,
2.15,2.5,1.52,3.31,-0.2,2.7,1.89,1.8,2.61,1.34,2.55,3.87,1.58,3.29,2.97,
2.33,2.6,2.65,2.32,1.23,2.22,2.83,1.52,1.85,1.84,1.8,1.75,1.96,1.31,0.47,
1.3,2.69,2.01,3.41,1.03,1.44,0.65,1.76,1.72,0.85,3.58,1.2,2.2),5,20)

# Standardize data:
S<-cov(rbind(x,y))
C<-diag(1/sqrt(diag(S)))
x2<-x %*% C
y2<-y %*% C

# Decide which matrices to use for the random projections:
matrixlist<-foreach(j = 1:B) %dopar%
{
   matrix(rnorm(k*p),p,k)		
}

# Compute test statistics:
T.raw<-projectionT2(x,y,n1,n2,p,B,k,matrixlist)
T.standardized<-projectionT2(x2,y2,n1,n2,p,B,k,matrixlist)

# Permutations:
z<-rbind(x,y)
z2<-rbind(x2,y2)
zsize<-n1+n2
	
rs<-data.frame(foreach(i = 1:B2) %dopar%
{
   matrixlist2<-foreach(j = 1:B) %dopar%
   {
      matrix(rnorm(k*p),p,k)		
   }
   x.rows<-sample(zsize,n1)
   x.new<-z[x.rows,]
   y.new<-z[-x.rows,]
   x2.new<-z2[x.rows,]
   y2.new<-z2[-x.rows,]
   c(raw=projectionT2(x.new,y.new,n1,n2,p,B,k,matrixlist2),
   stand=projectionT2(x2.new,y2.new,n1,n2,p,B,k,matrixlist2))
})

raw<-as.numeric(rs[1,])
stand<-as.numeric(rs[2,])

# p-value for raw data:
sum(raw>=as.numeric(T.raw))/B2

# p-value for standardized data:
sum(stand>=as.numeric(T.standardized))/B2
\end{verbatim}

\end{document}